\documentclass[runningheads]{llncs}
\usepackage{graphicx}
\usepackage{subcaption}
\usepackage[numbers,sort&compress]{natbib}
\usepackage{xcolor}

\usepackage{booktabs}

\begin{document}
%
\title{Moral Framing and Ideological Bias of News}
%
%
\author{Negar Mokhberian\inst{1} \and
Andr\'es Abeliuk\inst{1} \and
Patrick Cummings\inst{2} \and
Kristina Lerman \inst{1} }
\authorrunning{N. Mokhberian et al.}
%
\institute{Information Sciences Institute, University of Southern California, Marina del Rey, CA, USA.\\
\email{\{nmokhber,aabeliuk,lerman\}@isi.edu}\\
 \and
Aptima, Woburn, MA, USA\\
\email{pcummings@aptima.com}
}
\maketitle       
\begin{abstract}
News outlets are a primary source for many people to learn what is going on in the world. However, outlets with different political slants, when talking about the same news story, usually emphasize various aspects and choose their language framing differently. This framing implicitly shows their biases and also affects the reader's opinion and understanding. Therefore, understanding the framing in the news stories is fundamental for realizing what kind of view the writer is conveying with each news story.
In this paper, we describe methods for characterizing moral frames in the news. We capture the frames based on the Moral Foundation Theory. This theory is a psychological concept which explains how every kind of morality and opinion can be summarized and presented with five main dimensions. We propose an unsupervised method that extracts the framing Bias and the framing Intensity without any external framing annotations provided. We validate the performance on an annotated twitter dataset and then use it to quantify the framing bias and partisanship of news.

\keywords{Framing \and Bias \and Moral Foundation Theory \and News.}
\end{abstract}
\section{Introduction}
 A growing share of Americans receives vital information and news from online sources~\cite{gottfried2016news}. 
In recent years, however, the online information environment has grown increasingly polarized along political and ideological lines~\cite{prior2013media}.
Anecdotal evidence suggests that the shift to online news consumption has accelerated during the uncertainty and the `fog of war' created by the Covid19 pandemic\footnote{\url{https://www.wsj.com/articles/trump-mobilizes-the-white-house-to-tackle-coronavirus-but-adds-to-the-fog-of-war-11585934823}}, accompanied by growing ideological polarization, which colors how information about the pandemic is produced and consumed~\cite{van2020using}. These converging developments suggest a need for accurate methods to quantify the ideological bias of news sources. The resulting metrics could help the public become more responsible consumers of news by cutting through polarization.

Measuring ideological biases from text is a problem that has only recently attracted attention from the Natural Language Processing (NLP) community. 
Beyond identifying topics in the text, this research attempts to uncover how implicit biases manifest themselves through language and how to infer them from the text. Research has shown that language captures implicit associations that shape how people perceive the world and relationships within it, including gender and career representations that lead people to associate women with artistic careers and men scientific careers~\cite{caliskan2017semantics}.  
In cognitive linguistics and communication theory, the different perspectives and meanings that texts carry are captured by ``semantic frames''. When applied to news media, framing captures how news stories express meaning and cultural values and how they evoke an emotional reaction from readers. Media framing has been studied as a tool to influence the opinion of newsreaders \cite{liu-etal-2019-detecting}. 

Political scientists have characterized the framing of text using  Moral Foundations Theory (MFT)~\cite{haidt2004intuitive}. MFT defines five moral foundations which categorize the intuitive ethics and automatic emotional reactions to various social situations.  These foundations concern dislike for the suffering of others (care/harm), dislike of cheating (fairness/cheating), group loyalty (loyalty/betrayal), respect of authority and tradition (authority/subversion), and concerns with purity and contamination (sanctity/degradation). 
Across cultures, assessments of moral foundations reveal correlations between political ideology and the five foundations~\cite{graham2011mapping}, as well as interesting relationships between gender, culture, religion, and personality. For example, consistently and across studies,  liberals and conservatives draw upon these foundations to different degrees, measured via questionnaires~\cite{graham2009liberals}: liberals assign more weight to the Harm and Fairness foundations, whereas conservatives are more sensitive to the Ingroup, Authority, and Purity.

We use MFT as a generalizable framework to quantify the framing Bias in the news. The five moral foundations (care/harm, fairness/cheating, loyalty/betrayal, authority/subversion, sanctity/degradation) give a high-level understanding of the values promoted by news sources, which we use to quantify the biases inherent in how the news is framed by different sources. 

In this paper, we propose a framework to quantify the moral framing of texts. The framework leverages a large corpus of tweets annotated with their moral foundations~\cite{twitter35k}. 
We use sequence embeddings as features to train a classifier to predict the scores of text corresponding to the moral dimensions. Additionally, instead of using the embedding directly, we evaluate a low-dimensional feature representation of the text based on the Frame Axis approach~\cite{kwak2020frameaxis}. The Frame Axis method computes the Bias and Intensity of each moral frame based on alignments of the text relative to specific target words. 
We show that these simple features capture the moral frames implicit in the text, at least as well as much more complex (high-dimensional) representations. Finally, we show that moral frames help with the prediction of the partisanship of news based on the headlines. Our work demonstrates the feasibility of automatically classifying the moral framing and political partisanship of news.


\section{Background} 
\paragraph{\textbf{Moral Foundations Theory.}}
Moral Foundation Theory (MFT) has been first introduced by Haidt and Joseph to explain moral differences across cultures \cite{haidt2004intuitive}. The theory introduces five basic moral foundations which are the basis of many intuitive and cultural human psychological values. These five dimensions consist of Care/Harm, Fairness/Cheating, Loyalty/Betrayal, Authority/Subversion, and Purity/Degradation.
Later, Graham and Haidt showed that Liberals and Conservatives have substantial variation in their moral concerns across these five Moral Foundations~\cite{graham2009liberals}.
This triggered future works analyzing the political rhetoric based on MFT and characterizing the political ideologies using the five Moral Foundations.

Other than the works in the psychology domain, there are other works considering MFT in computational linguistic approaches. They are mostly using the lexical resource Moral Foundation Dictionary (MFD) \cite{graham2009liberals}. This dictionary consists of words regarding virtues and vices of the five Moral Foundation dimensions and a sixth dimension regarding general morality terms. Studies rely on the usage of words with respect to the Moral Foundation dictionary terms. For example, to analyze temporal changes in the frequency of MFT terms in English books for the years 1900 to 2007~\cite{wheeler2019twentieth}, or to demonstrate changes in the Moral Foundation language regarding the word 'gay' in the US political Senate speeches from 1988 to 2012~\cite{garten2016morality}. 
The study showed that republicans were significantly using more Purity words than Democrats. 

Recently, novel methods have been developed to recognize if a text is relevant to any of the MFT dimensions. This problem has been formulated as a classification problem on tweets in \cite{garten2016morality, johnson-goldwasser-2018-classification} and on news in \cite{fulgoni-etal-2016-empirical}.

\paragraph{\textbf{NLP}} 
Some of the previous work on analyzing MFT in text corpora have relied on the word counts \cite{fulgoni-etal-2016-empirical}.
Other studies \cite{garten2016morality, Brendan} have used features based on word embeddings \cite{pennington2014glove} or sequence embeddings \cite{devlin-etal-2019-bert} to obtain more robust models and better performance. 
Recent work has proposed methods for analyzing rhetorical frames in text. In SemAxis~\cite{an-etal-2018-semaxis}, the authors introduce semantic axis which are word-level domain semantics structured on word antonym pairs. The similarity of a word with respect to different predefined antonymous axes can capture the semantics of the word in various contexts. (e.g., the word 'soft' can be a negative word in the context of sports and positive in the context of toys). The word and the semantic axis are represented in the same representation vector space trained on a corpora. The authors of SemAxis later introduced the concept ``Frame Axis'' which is a method of characterizing the framing of a text by identifying the most relevant semantic axis \cite{kwak2020frameaxis}. Similarly, studies have explored using groups of words with opposite meanings to define semantic dimensions to improve the interpretability of text representations~\cite{10.1145/3366423.3380227}.




\section{Methods}
In this section, we describe our computational framework to measure the framing bias of news sources based on the text of their headlines. The framework is composed of two tasks:
(1) An unsupervised method to learn a low-dimensional representation characterizing the text with respect to a set of target words; (2) A supervised classification task using Twitter data with human-annotated MF framing, with contextualized language representation models.

In the first task, we represent the text with scores according to the moral dimensions
, computing framing Bias and Intensity scores for each MF using the Frame Axis \cite{kwak2020frameaxis} method. 
This approach projects the words on micro-frame dimensions characterized by two sets of opposing words.
These MF framing scores capture the ideological slant of the news source and can be used as features for predicting partisanship. Details are given in Section~\ref{sec:unsupervised}.

In the second task, we leverage the annotated Twitter dataset \cite{twitter35k} to develop a supervised model to classify the MFT related micro-frames in the news headlines (see Section~\ref{sec:supervised}). 
In section~\ref{sec:twitterValidating}, we validate the accuracy of different latent representations for the text on the annotated Twitter data set. 

Finally, in section~\ref{sec:classifypartisan} we present a case study of our framework analyzing the moral framing differences between Liberal and Conservative media in USA news sources. 
\subsection{Data}
\paragraph{\textbf{News Articles}}
We use ``All the News'' dataset from Kaggle\footnote{\url{https://www.kaggle.com/snapcrack/all-the-news}}. The data primarily falls between the years 2016 and July 2017. The news sources include the New York Times, Breitbart, CNN, Business Insider, the Atlantic, Fox News, Buzzfeed News, National Review, New York Post, the Guardian, NPR, Reuters, Vox, and the Washington Post.
For each news story, we have worked with its headline and publication (the news outlet). Also, we have looked up the political leaning of each source from allsides\footnote{\url{https://www.allsides.com/media-bias/media-bias-ratings}} website and have added a column indicating the political side of each news. We have eliminated the news stories from central sources and only kept the liberal and conservative leaning sources. We have narrowed down the headlines to the ones related to immigration and elections topics. We did this by checking the headlines to include some hand-picked words regarding each topic. Among several topics, we chose these two because the frequency of articles falling in these two topics was larger.  After all the above steps, the data consisted of 3242 news articles about immigration and 29345 regarding elections.

\paragraph{\textbf{Annotated MF Data.}}
We use the annotated twitter dataset~\cite{twitter35k} to train and test classifiers for each MF micro-frame. Several trained human annotators determined which of the MF virtues or vices are most relevant for each tweet, or if there are no moral concepts related to that tweet. In total there are 11 dimensions for each tweet (virtues and vices of the five MFT dimensions and also non-moral dimension). Each annotator can assign more than one MF dimension to a single tweet. To aggregate the votes for a single tweet, we have assigned 1 to the dimensions having at least two votes and 0 to the dimensions that have less than two votes. There were 35k tweet ids provided in this dataset and at least three annotators per tweet.

\begin{table}[t!]
  \centering
  \scriptsize{
  \begin{tabular}{c|c|c|c|c|c}
  \hline
\textbf{Care}	&	\textbf{Fairness}	&	\textbf{Ingroup}	&	\textbf{Authority}	&	\textbf{Purity}	&	\textbf{General Morality}	\\ \hline
\multicolumn{6}{c}{\textit{Virtues}}\\ \hline
care	&	fair	&	ally	&	abide	&	austerity	&	blameless	\\
benefit	&	balance	&	cadre	&	allegiance	&	celibate	&	canon	\\
amity	&	constant	&	clique	&	authority	&	chaste	&	character	\\
caring	&	egalitarian	&	cohort	&	class	&	church	&	commendable	\\
compassion	&	equable	&	collective	&	command	&	clean	&	correct	\\
empath	&	equal	&	communal	&	compliant	&	decent	&	decent	\\
guard	&	equity	&	community	&	control	&	holy	&	doctrine	\\
peace	&	fairminded	&	comrade	&	defer	&	immaculate	&	ethics	\\
protect	&	honest	&	devote	&	father 	&	innocent	&	exemplary	\\
safe	&	fair	&	familial	&	hierarchy	&	modest	&	good	\\
secure	&	fairly	&	families	&	duty	&	pious	&	goodness	\\
shelter	&	impartial	&	family	&	honor	&	pristine	&	honest	\\
shield	&	justice	&	fellow	&	law	&	pure 	&	legal	\\
sympathy	&	tolerant	&	group	&	leader	&	sacred	&	integrity	\\ \hline
\multicolumn{6}{c}{\textit{Vices}}\\ \hline					
abuse	&	bias	&	deceive	&	agitate	&	adultery	&	bad	\\
annihilate	&	bigotry	&	enemy	&	alienate	&	blemish	&	evil	\\
attack	&	discrimination	&	foreign	&	defector	&	contagious	&	immoral	\\
brutal	&	dishonest	&	immigrant	&	defiant	&	debase	&	indecent	\\
cruelty	&	exclusion	&	imposter	&	defy	&	debauchery	&	offend	\\
crush	&	favoritism	&	individual	&	denounce	&	defile	&	offensive	\\
damage	&	inequitable	&	jilt	&	disobey	&	desecrate	&	transgress	\\
destroy	&	injustice	&	miscreant	&	disrespect	&	dirt	&	wicked	\\
detriment	&	preference	&	renegade	&	dissent	&	disease	&	wretched	\\
endanger	&	prejudice	&	sequester	&	dissident	&	disgust	&	wrong	\\
fight	&	segregation	&	spy	&	illegal	&	exploitation	&		\\
harm	&	unequal	&	terrorist	&	insubordinate	&	filth	&		\\
hurt	&	unfair	&		&	insurgent	&	gross	&		\\
kill	&	unjust	&		&	obstruct	&	impiety	&		\\ \hline
\end{tabular}
}
\caption{Some of the positive and negative words (virtues and vices) associated with the five dimensions of the Moral Foundations Theory and general morality. }
  \label{tab:mft-words}
\end{table}

\subsection{Quantifying Moral Frames with Frame Axis} 
\label{sec:unsupervised}
For the unsupervised method, we quantify the strength of the moral framing of text along the dimensions of MFT using the Frame Axis approach~\cite{kwak2020frameaxis}.
Frame Axis proposes two measures---\textit{Intensity} and \textit{Bias}---to capture the document-level framing based on the word contributions \cite{kwak2020frameaxis}. Intensity and Bias for a text are calculated as the weighted average of mapping of its words towards the desired semantic axis.

Each semantic axis (also called micro-frame) builds on a set of antonyms, i.e., words with opposite meaning \cite{an-etal-2018-semaxis}. In our case, we choose the vices and virtues of the Moral Foundations as opposites of a word axis, e.g. Care terms vs Harm terms from the MFT. Some of these words are shown in Table~\ref{tab:mft-words}. For each moral foundation (MF) dimension, the axis is calculated by subtracting the average vector of the embeddings of positive words (virtues) and the average vector of the embeddings of negative words (vices) of that MF dimension. Formally, let $m$ be one of the MF dimensions (e.g. Care) and $V_{m}^+$ denote the set of embedding vectors of virtue words (e.g. vectors for Care words) and $V_{m}^-$ denote to set of vectors of vice words (e.g. vector for Harm words), then the semantic axis corresponding to this MF dimension is:
\begin{equation}
  A_{m} = mean(V_{m}^+)-mean(V_{m}^-)
\end{equation}

For the computations of this part, the embeddings of words are obtained from the pretrained GloVe model \cite{pennington2014glove} called ``Common Crawl" which includes 2.2M vocab\footnote{\url{https://nlp.stanford.edu/projects/glove/}}. 
Following \cite{kwak2020frameaxis}, we define the \textit{framing Bias} $B_m^{D}$ of a document $D$ 
along a semantic axis $m$ as:
\begin{equation}
B_m^{D}=\frac{\sum_{d\in D}f_{d}\ s(A_{m},d)}{\sum_{d\in D}f_{d}},
\label{eq:bias}
\end{equation}
where a document $D=\left\{ d_{1},\ldots,d_{n}\right\} $ is defined as a set of embeddings of its words;
$s(A_{m},d)$ is the cosine similarity between the semantic axis $A_{m}$ and word $d$; $f_{d}$ is the frequency of word $d$ in the document. In other words, the Bias of a text with regard to a moral foundation axis $m$ is the weighted average of the cosine similarity of its words with that axis. 
Notice that, if the embeddings represent sentences then there are no repetitions, i.e., $f_{d}=1$. The absolute value of the Bias captures the relevance of the document to the moral dimension, while the sign of the similarity captures a bias toward one of the poles in the moral dimension. The positive sign of Bias will shows the document is aligned with the positive pole of the Axis and negative sign shows the opposite.

Second, we use \textit{framing Intensity} on a moral dimension to capture how heavily that moral dimension appears in the document with respect to the background distribution: 
\begin{equation}
I_{m}^{D}=\frac{\sum_{d\in D}f_{d}\left(s(A_{m},d)-B_{m}^{T}\right)^{2}}{\sum_{d\in D}f_{d}},
\end{equation}
where $B_{m}^{T}$ is the baseline framing Bias of the entire text
corpus $T$ on a moral dimension $m$. Intensity doesn't reveal information about the polarization. However, in situations that both poles of an axis actively appear in a text, the positive and negative terms will cancel out each other, and the document wouldn't have a significant Bias toward any pole of that axis, but Intensity will show the relevance to that axis.

As a result, each document can be represented by 12 dimensions, each representing the Bias and Intensity scores for each of the 6 MF dimensions in Table~\ref{tab:mft-words}.

\subsection{Classifying Moral Frames from Text}
\label{sec:supervised}

In the supervised approach, we leverage the corpus of tweets, manually annotated with their moral foundations, as train data to learn a classifier model on MF frames from text. 

We trained a binary classifier on the twitter dataset to learn the relevance of the Moral Foundations.
Specifically, each MF is considered as a label that can get values 0 or 1, and for each MF we train a binary classifier. A Logistic Regression classifier is used for this part.

For creating text features, we use a contextualized sequence encoding method known as Bidirectional Encoder Representations from Transformers (BERT) \cite{devlin-etal-2019-bert} to obtain the embeddings for each tweet. We encode each tweet in a 768-dimensional encoding using a pre-trained BERT model. 

In the inference phase, we convert the text in the test dataset to the same feature space, and each classifier gives a likelihood showing how much the given text is related to the corresponding moral foundation. 

\begin{table}[t!]
\begin{center}
\small{
\begin{tabular}{ c|c|c|c|c|c|c} 
\hline
\textbf{Moral Foundation }    & \textbf{Precision} & \textbf{Recall} & \textbf{F1}  & \textbf{Accuracy}  & \textbf{Baseline F1} & \textbf{Baseline Acc.} \\ \hline
& \multicolumn{6}{c}{\textit{BERT Embedding features}} \\ \hline
AVG    & 0.771   & 0.822 & 0.775 & 0.822 & 0.705 & 0.705     \\ \hline
authority & 0.808   & 0.875 & 0.817 & 0.875 & 0.778 & 0.776     \\ \hline
fairness & 0.662   & 0.774 & 0.681 & 0.774 & 0.655 & 0.655     \\ \hline
harm   & 0.746   & 0.768 & 0.734 & 0.768 & 0.613 & 0.613     \\ \hline
ingroup  & 0.802   & 0.873 & 0.816 & 0.873 & 0.779 & 0.781     \\ \hline
purity  & 0.910   & 0.935 & 0.908 & 0.935 & 0.527 & 0.527    \\ 
 \hline
morality & 0.698   & 0.705 & 0.694 & 0.705 & 0.879 & 0.879     \\ \hline \hline
& \multicolumn{6}{c}{\textit{Frame Axis features}} \\ \hline
AVG    & 0.787 & 0.818 & 0.773 & 0.818 & 0.705 & 0.705 \\ \hline
authority & 0.883 & 0.888 & 0.851 & 0.888 & 0.778 & 0.776 \\ \hline
fairness & 0.770 & 0.795 & 0.743 & 0.795 & 0.655 & 0.655 \\ \hline
harm   & 0.694 & 0.740 & 0.666 & 0.740 & 0.613 & 0.613 \\ \hline
ingroup  & 0.799 & 0.873 & 0.816 & 0.873 & 0.779 & 0.781 \\ \hline
purity  & 0.898 & 0.933 & 0.907 & 0.933 & 0.879 & 0.879 \\ \hline
morality & 0.676 & 0.683 & 0.655 & 0.683 & 0.527 & 0.527 \\ \hline
\end{tabular}
}
\end{center}
\caption{Evaluation of moral foundation classifiers on annotated tweets.}
\label{table:twitterClassifier}
\end{table}

\section{Results}

\subsection{Measuring Moral Framing from Text}
\label{sec:twitterValidating}
First, we evaluate the ability of the two latent representations to measure MF frames from the text on the annotated twitter dataset. We compare the classification performance of the tweet BERT embeddings to the Frame Axis features on the annotated tweets. 
We run the classification task repeatedly on random 0.75/0.25 train/test splits. Classification results in Table~\ref{table:twitterClassifier} show that both approaches dramatically outperform the baseline. The baseline predicts each moral foundation according to its frequency distribution in the training set. They also outperform the method from \cite{twitter35k}, which reported $F1<0.5$ on a subset of the twitter dataset. 
Remarkably, Frame Axis achieves comparable performance to the embedding-based approach using only two features (Bias and Intensity) for each moral dimension.

\begin{figure}[t!]
 \begin{subfigure}{0.333\textwidth}
  \includegraphics[width=\linewidth]{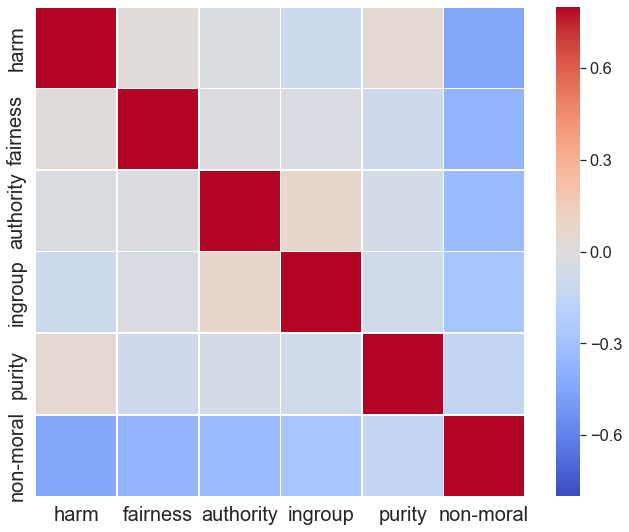}
  \caption{Original Twitter Annotations} \label{fig:1a}
 \end{subfigure}%
 \begin{subfigure}{0.333\textwidth}
  \includegraphics[width=\linewidth]{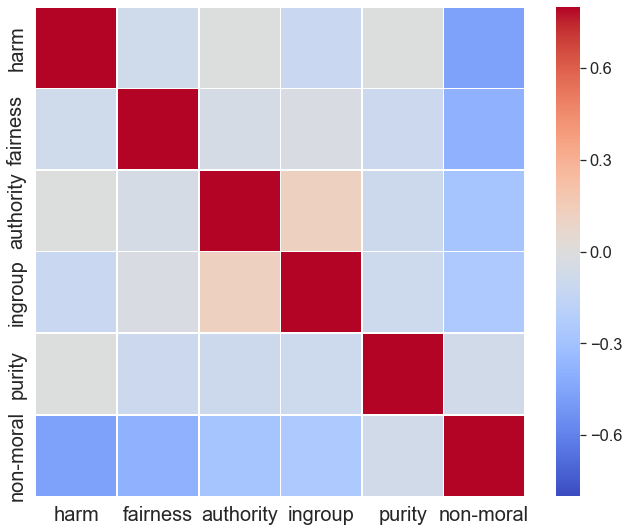}
  \caption{Twitter learned frame scores} \label{fig:1b}
 \end{subfigure}%
 \begin{subfigure}{0.333\textwidth}
  \includegraphics[width=\linewidth]{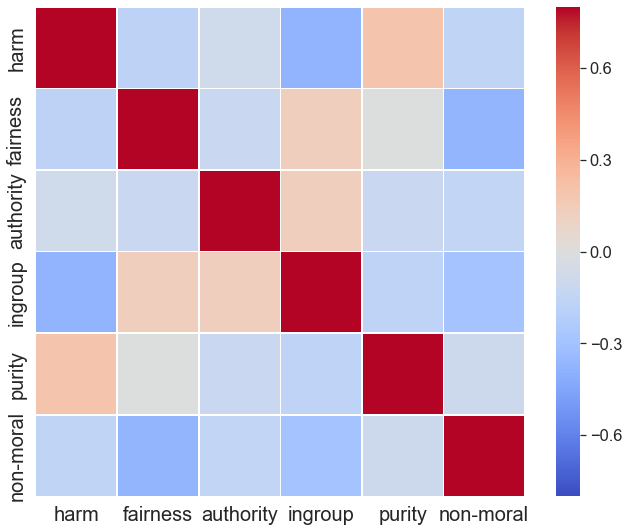}
  \caption{News learned frame scores} \label{fig:1c}
 \end{subfigure}
\caption{The correlation (a) among the count of annotators selected a MF for each tweet (b) among the MF frame likelihoods learned from the supervised method on the tweets and (c) on the news headlines.} \label{fig:corrplots}
\end{figure}
\subsection{Relationships between Moral Framings}
In this section, we explore the empirical correlations among the different MF frames learned from the data. 
We take inspiration from previous work studying correlations using surveys based on the Moral Foundations Questionnaire~\cite{graham2009liberals}.
For this purpose, we compute the correlation matrix using two approaches: 1 ) based on the original hand-annotated MF frames on the Twitter dataset; and 2) based on the inferred frames that we obtained from our supervised method.  With the latter method, we can infer the MF frames correlation matrix for the Twitter data and the news articles corpus. To evaluate the method, we compare the inferred and hand-annotated correlation of MF frames. Finally, we explore the correlations learned from the news headlines.

The annotated Twitter dataset contains the number of human annotators who have selected each tweet as relevant to a MF frame. We can compare the correlation matrix of these expert annotations to the likelihoods calculated by the classification task for each MF frame. Figure~\ref{fig:corrplots} shows that the correlations between moral foundation dimensions on the raw annotations (Fig.~\ref{fig:corrplots}a) are very similar to the correlations between MFs likelihoods predicted by the supervised classifier on the same Twitter data (Fig.~\ref{fig:corrplots}b). 
Even though we are using distinct classifiers for each MF and our classifier does not see different labels at the same time, still the correlations in (Fig.~\ref{fig:corrplots}b) are comparable to (Fig.~\ref{fig:corrplots}a), showing that the automatically detected frames are consistent with human judgment.
For example, we see that similar to the original Twitter dataset, the correlation of non-moral and other moral foundations stay negative, which makes sense because if a text is not showing any MF frame, then the highest score for it must be the non-moral label and all the other frame scores must be low.
However, when we use the supervised classifier to compute MF frame scores on the news headlines, we see different correlations between the moral dimensions (Fig.~\ref{fig:corrplots}c). Here, Purity is more correlated with the Care/Harm dimension, and Ingroup is more correlated with Fairness and Authority than in Twitter data. 

\begin{table}[t!]
\begin{center}
\begin{tabular}{|l|l|l|l|l|}
\hline
& \multicolumn{2}{l|}{Immigration} & \multicolumn{2}{l|}{Election} \\ \hline
features/approach & F1            & Accuracy         & F1           & Accuracy        \\ \hline
Baseline          & 0.50          & 0.51             & 0.50         & 0.51            \\ \hline
MF Likelihoods     & 0.61          & 0.63             & 0.60         & 0.61            \\ \hline
Frame Axis    & 0.68          & 0.69             & 0.66         & 0.66            \\ \hline
MF Likelihoods + Frame Axis           & 0.69            & 0.70               & 0.67           & 0.67              \\ \hline
\end{tabular}
\end{center}
\caption{The results on classifying partisanship of headlines with the likelihoods calculated from the Logistic Regression classifier. The classifier was trained with three different input features: 1) the tweet BERT embeddings; 2)the Bias and Intensity scores from the Frame Axis approach; 3) the combinations of the two sets of described features. The baseline predicts the result based on the training set distributions.}
\label{table:partisan}
\end{table}
\subsection{Moral Framing and Partisanship of News}
\label{sec:classifypartisan}
\subsubsection{Predicting Partisanship of News}
Using moral frames as features, we classify the partisanship of news headlines. 
We have chosen Immigration and Elections as two categories of news for experimenting. 
Our dependent variable is the partisanship of the news source (Liberal or Conservative), and we use Bias and Intensity scores for each moral foundation and the likelihoods obtained from the classifier trained on the tweets as features to test for systematic differences in the moral framing of news. 
We test whether moral frame scores can distinguish between ideologies of news sources from different political sides. The Bias and Intensity are unsupervised measures since for calculating those, no annotations are needed. Another feature set for representing MF framing can be obtained providing the news headlines as test data to the classifier previously trained on the annotated twitter data. This supervised classifier gives likelihoods corresponding to each of the MF dimensions which we use to classify the partisanship. 
Table~\ref{table:partisan} shows F1 and accuracy of classifying the partisanship based on supervised MF likelihoods and unsupervised Frame Axis scores used as features. The row `combine' denotes concatenating these two feature sets.
The baseline is a simple classifier that learns the distribution of partisanship from the training data and uses it to make predictions for the test data. 
Features based on moral frames outperform the baseline by a wide margin. 


\begin{figure}[t!]
\centering
    \includegraphics[width=\linewidth]{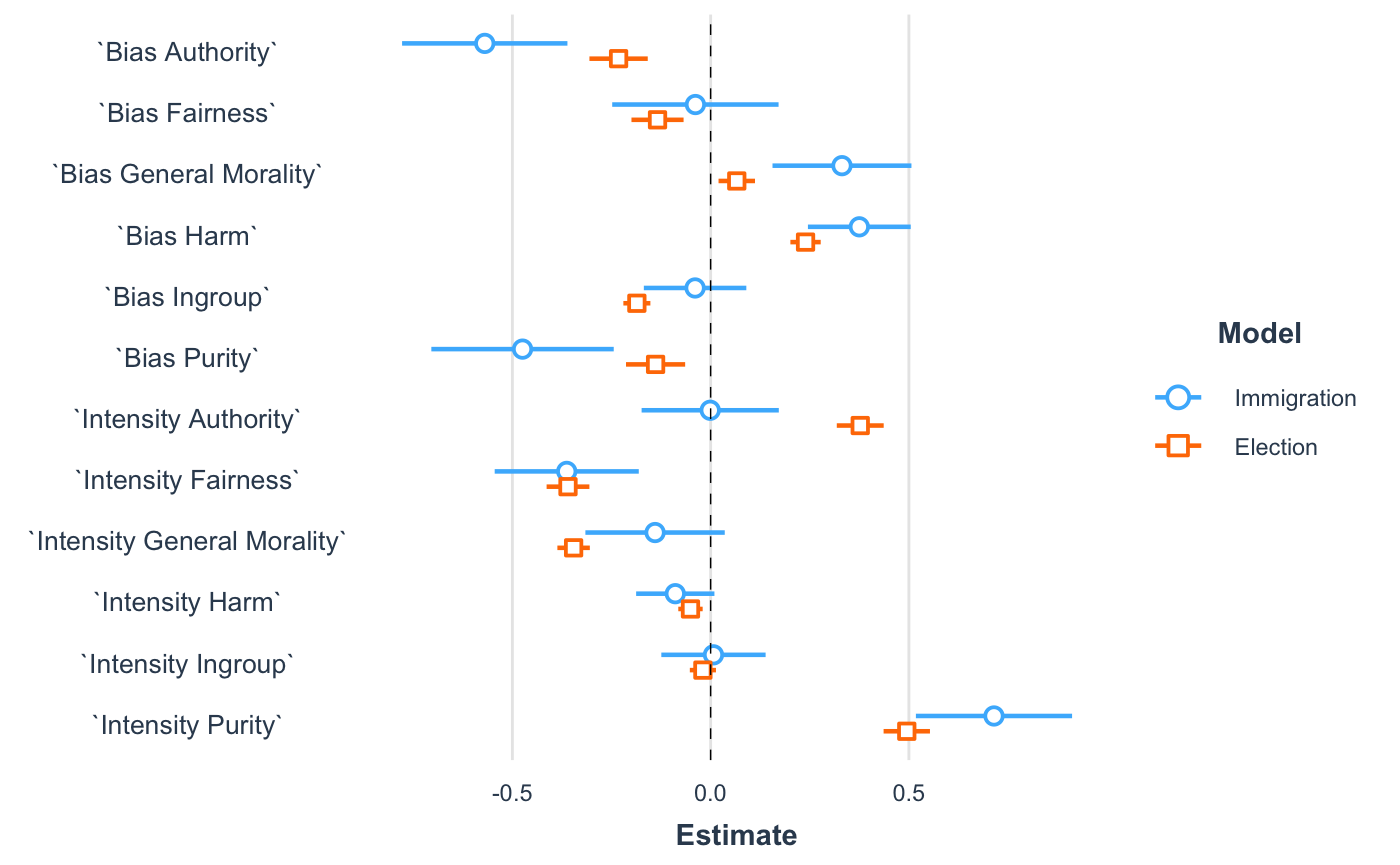}
    \caption{Coefficients of Logistic Regression classification task using moral frames to predict liberal partisanship. The coefficients whose 0.95 confidence intervals exclude 0 are significant. The binary classifier target label is the partisanship with Liberal as 1 and Conservative as 0.} 
    \label{tbl:log_results}
\end{figure}
\subsubsection{Moral Framing of News}
Next, to quantify systematic differences in moral framing between liberal and conservative news articles, we inspect the model coefficients learned by the partisanship classifier.  One key aspect of using the Frame Axis approach is that the model coefficients are straightforward to interpret (see Section~\ref{sec:unsupervised}) and have a competitive performance.
Figure~\ref{tbl:log_results} reports the Frame Axis coefficients and their $0.95$ confidence intervals for Intensity and Bias features for each moral foundation.
Since in our setting, we coded the label for Conservative partisanship as 0 and Liberal partisanship as 1, the interpretation of a positive coefficient is that all else being equal, Liberal news articles are more likely than Conservative articles to exhibit the corresponding attribute.
We highlight the following findings: 
\begin{itemize}
\item The sign of the coefficients are consistent across the two topics analyzed and show significant differences across partisanship for most moral foundations.
\item Purity/Degradation foundation. The high and positive Intensity coefficient indicates that liberal media stress the purity/degradation foundation more strongly than conservative media when reporting news about immigration and elections. The negative bias coefficient implies that liberal media emphasize more 
the recognition of vices being violated regarding dirtiness, unholiness, and impurity, whereas conservative media tend to emphasize more the virtues like austerity, sacred, and pure. 
\item Authority/Subversion foundation.
The Intensity of Authority is insignificant for the Immigration topic. However, it is substantial for the Election topic. This shows liberals and conservatives hold significantly different views on Authority in the Election topic. Liberal media, framing their articles  with more attention to the Authority/Subversion foundation, contrary to the withheld consensus~\cite{graham2009liberals}. The negative 
Bias suggests a stronger framing on vices words that describe rebellion by Liberal media.
\item Care/Harm foundation.  
The negative Intensity coefficient suggests that conservatives tend to endorse more strongly this foundation, especially in election articles.  
The positive Bias shows that the framing of conservative media, compared to liberal media, has a higher emphasis on the vice side of the care/harm moral dimension, which condemns malice, abuse, and inflicting suffering.
 
\item The large positive coefficient of Bias for General Morality suggests that liberal media tends to frame their news about immigration with a more positive moralistic view, indicative of normative judgments (e.g., good, moral, noble). While Conservative media use a more negative moral judgment based on words like bad, incorrect, or offensive.
\end{itemize}


\section{Conclusion}
Recent research puts in evidence a change in partisanship among the general electorate, where the number of issues with partisan conflict has increased~\cite{brewer2005rise}. 
Studies suggest a link between partisan media exposure and polarization~\cite{sunstein2009going}, driven by motivated reasoning to explain why partisan media polarizes viewers~\cite{levendusky2013partisan}.
An alternative approach emphasizes the impact on how the news media frame political discourse. In their seminal study, Kahneman and Tversky, demonstrate the impact of framing in human decisions~\cite{tversky1989rational}. Since then, there has been vast experimental evidence on how moral framing can influence attitudes towards polarizing issues like climate change~\cite{hurst2020messaging}.

In this paper, we focused on developing computational methods to detect moral framing Biases in news media using NLP techniques. Inspired by the work done on Frame Axis~\cite{kwak2020frameaxis}, identifying meaningful micro-frames from antonym word pairs, coupled with the principled moral foundation theory, we choose the vices and virtues of the five Moral Foundations as polar opposites of a micro-Frame Axis. We proceed to first, validate our approaches on an annotated Twitter dataset, and second, study the moral framing Bias on partisan news articles related to immigration and elections topics. Our findings reveal systematic differences across liberal and conservative media.  It supports the correlations between political ideology and the five foundations which have been shown in the empirical evidence based on surveys~\cite{graham2009liberals}. In particular, our results suggest that rhetoric on Purity is different between liberal media compared to conservative media, where liberals
tend to use more rhetoric towards the violations of the Purity moral foundation. This observation, supporting the growing evidence that Purity is among the most differentiating moral dimension between conservatives and liberals~\cite{dehghani2016purity}.\\

Our work has some limitations that should be noted. First, our methods rely on the dataset, and changing the dataset might change the results. We have demonstrated our methods on two different topics immigration and elections.
Second, in the calculation of Bias, as defined in equation (\ref{eq:bias}), the negations in the sentence which can change the polarization are not considered. e.g., ``This is not fair" would have a positive Bias in Fairness micro-frame because the presence of the word ``not" is not considered in the definition of Bias. However, this problem does not appear in Intensity because that measures the frequency of usage of words from both poles of each axis.
Lastly, even though there is a twitter dataset annotated with moral foundations, there is no annotated dataset of news articles. In the supervised method described in section \ref{sec:supervised}, we have trained the classifier on the labeled tweets and then predicted the moral foundations of news headlines using that model. 

As future work, we plan to study different topics and compare framing across different news sources. Most of the studies have paid attention to liberal and conservative news sources. An interesting question is how central news sources MF framing would be different from the polarized ones?
Another path to continue this work will be leveraging the BERT text encodings by fine tuning it according to our MF frame classification task. 

\section{Acknowledgements}
This work was conducted in connection with Contracts \#W56KGU-19-C-0004 with the U.S. Army Combat Capabilities Development Command C5ISR Center. The views, opinions, and findings contained in this document are those of the authors and should not be construed as an official position of the United States Army.

%
%
\bibliographystyle{splncs04} 
\bibliography{references}

\end{document}